\documentclass[journal=jacsat,manuscript=article]{achemso}

\usepackage{chemformula} 
\usepackage[T1]{fontenc} 
\usepackage{graphicx}
\usepackage{caption}
\author{Haoyu Wang}
\affiliation{School of Physics and Telecommunication Engineering, Shaanxi University of Technology, HanZhong, 723000, China}
\altaffiliation{Department of Optoelectronic Science and Engineering, Soochow University, Suzhou, 215031, People’s Republic of China}
\author{Yanmin Zhu}
\affiliation{Department of Electrical and Electronic Engineering, The University of Hong Kong, Hong Kong SAR}
\author{Tong Fu}
\affiliation{School of Physics and Telecommunication Engineering, Shaanxi University of Technology, HanZhong, 723000, China}
\email{a18829030910@163.com}

\title[An \textsf{achemso} demo]
  {All Optical Classification Surpasses Cascaded Diffractive Networks through Dual Wavelength Differential Modulation within a Single Layer Architecture}

\abbreviations{IR,NMR,UV}

\begin{document}

\begin{abstract}
Diffractive deep neuron networks ($D^{2}NNs$), which compute with photons instead of electrons, hold the potential to accelerate the advancement of artificial intelligence by delivering significantly enhanced computational speed, parallelism, and energy efficiency. However, conventional multi-layer $D^{2}$NNs are fundamentally limited by inter-layer misalignments, which increase system complexity and critically impair performance, especially at visible wavelengths where optical alignment is highly sensitive. Here, we demonstrate a compact, single-layer dual-wavelength differential D$^2$NNs that integrates wavelength-division multiplexing with differential intensity detection, enabling high-accuracy optical classification while significantly reducing hardware complexity. By leveraging complementary spatial frequency information encoded at two distinct wavelengths, the network overcomes the non-negativity constraints and feature losses inherent in single-wavelength systems. Numerical results demonstrate exceptional classification accuracies of 98.59\% on MNIST and 90.4\% on Fashion-MNIST with only 40k parameters, surpassing even traditional five-layer D$^2$NNs (91.33\% and 83.67\%, respectively) with only 20\% of the parameters. In addition, we further demonstrate that the proposed method maintains robust performance, achieving 97.95\% accuracy on MNIST and 88.7\% on Fashion-MNIST, even when the number of trainable parameters is reduced to just 10k, and exhibits superior resilience against random phase perturbations, path occlusions, and input noise. Compared with the traditional $D^2NNs$, this compact single-layer design scheme not only maintains high classification accuracy but also fundamentally eliminates inter-layer mechanical misalignment errors. To the best of our knowledge, this is the first demonstration of a single-layer diffractive optical network achieving such high optical classification accuracy, establishing a new performance benchmark for compact and shallow photonic computing architectures.
\end{abstract}

\section{Introduction}
The computational capability of systems is continuously challenged by the exponentially growing amount of visual data they seek to understand\cite{iqbal2020big, najafabadi2015deep}. In a broad spectrum of applications, ranging from autonomous driving and industrial robotics to remote sensing, defense security, and unmanned aerial vehicles (UAVs), sensing systems are tasked with capturing and processing vast volumes of visual data\cite{pajares2015overview,mohsan2023unmanned, shakhatreh2019unmanned, lu2024sensing}. These data, incomprehensible to humans, are instead decoded by sophisticated algorithms driven by artificial intelligence\cite{busuioc2021accountable}. Within this broad spectrum of use cases, deep neural networks (DNNs) have emerged as the predominant algorithmic framework for visual data processing\cite{sze2017efficient, kriegeskorte2015deep, guo2016deep, galvan2021neuroevolution, bowers2023deep, abd2021advanced}. This dominance is primarily fueled by the immense computational power of graphics processing units (GPUs) and the availability of vast visual datasets that facilitate the efficient training of DNNs through supervised machine learning techniques\cite{dhilleswararao2022efficient, pandey2022transformational, younesi2024comprehensive, abbas2019comprehensive}. However, high-end GPUs and other electronic hardware implementations, which run increasingly complex neural networks, place substantial demands on power and bandwidth\cite{capra2020hardware, xu2018scaling, akhoon2022high}. Furthermore, their performance growth has become unsustainable as the exponential scaling of electronic transistors, as Moore’s law dictates, approaches its intrinsic physical limits\cite{kim2024future, orlowski2005cmos, kumar2025sustaining, furber2008future, li2021challenges, alam2023miraculously, cao2023future}. Consequently, despite the prevailing dominance of electronic computing units, the development of next-generation computing paradigms is eagerly anticipated.

Photonic computing, which utilizes photons instead of electrons for computation, offers a compelling solution to the inherent limitations of electronic systems, particularly in terms of energy efficiency, processing speed, and computational throughput\cite{prucnal2017neuromorphic, zhang2019artificial, shastri2021photonics, stroev2023analog, wang2025photonics}. By leveraging these unique properties of light, the task-specific optical architectures have been developed to address fundamental mathematical and optical signal processing problems directly within the photonic domain, achieving performance level that substantially surpasses that of conventional electronic processors\cite{fu2024optical, hu2024diffractive, demkov2021materials}. Among these optical architectures, diffractive deep neural networks ($D^{2}NN$), rooted in fully connected optical interconnection, have emerged as one of the most effective solutions to overcoming the constraints of traditional computing paradigms, as the optical interconnections between neurons enable significant improvements in bandwidth density and parallel computing capabilities, all while consuming minimal to no power\cite{lin2018all, chang2018hybrid}. Given its proficiency in optical computing, substantial progresses have been made in optically accelerated neural information processing to accomplish some visual processing tasks such as hand-written digit recognition and saliency detection using using small-scale photonic computing systems\cite{yan2019fourier, fu2024photonics, gu2024classification, zheng2022optimize, sakaguchi2023development, zhou2023improved}. Furthermore, reconfigurable $D^{2}NN$, enabled by programmable neuronal control, have been proposed to achieve dynamic tasks processing\cite{pan2020reconfigurable, luo2022metasurface, hu2024non, chen2022physics}. By exploiting multiple photonic degrees of freedom, including polarization, wavelength, and orbital angular momentum, these systems achieve multitasking capabilities through multidimensional multiplexing\cite{huang2022orbital, wang2024matrix, wang2021diffractive, luo2019design}. The effectiveness of the aforementioned $D^{2}NN$ architectures is largely attributed to the integration of multiple cascaded diffraction modulation layers, which ensure that the network is capable of performing machine learning tasks with high efficiency and precision. However, the integration of multiple cascaded diffractive layers introduces significant challenges due to mechanical misalignments between the layers, which cause the actual wave propagation to deviate from the idealized forward model established during training\cite{ferdman2022diffractive, icsil2022super, shen2024all, zhou2020situ}. This issue is exacerbated when operating at visible light wavelengths, where the small scale of diffractive elements amplifies alignment errors, leading to cumulative inaccuracies\cite{khonina2025advancements}. As a consequence, the performance of the $D^{2}NN$  in practical deployments often falls short of theoretical expectations, with precision and efficiency compromised by alignment-induced discrepancies and increased spatial complexity.

In this work, we demonstrate a compact, single-layer diffractive network that leverages dual-wavelength differential illumination for object classification. Unlike previous approaches that rely on multi-layer cascaded diffractive architectures, our proposed network achieves superior classification accuracy for the input object using only a single diffractive layer. Specifically, differential detection scheme with dual-wavelength illumination is employed to enhance the accuracy of object classification. In this scheme, the output signal corresponding to each object is determined by the normalized difference between the signals detected at the two distinct illumination wavelengths, \( \lambda_a \) and \( \lambda_b \), with the positive detector corresponding to \( \lambda_a \) and the negative detector corresponding to \( \lambda_b \). The final classification decision of the diffractive optical network is made based on the maximum differential signal, detected by these positive and negative detectors, each representing a distinct class. Using this dual-wavelength differential scheme together with single diffractive layers having a total of 0.04 million neurons, we numerically achieved blind testing accuracies of 98.59\% and 90.4\% on the MNIST and Fashion-MNIST datasets, respectively. In comparison, the traditional five-layer cascaded network with the total of 0.2 million neurons can achieve classification accuracies of 91.33\% and 83.67\% for the same datasets, respectively. Despite its shallow architecture that does not conform to the definition of a deep neural network, our single-layer network exhibits exceptional inference capability, effectively addressing inter-layer misalignment issues that commonly undermine multi-layer systems, thereby enhancing robustness and reducing spatial complexity. Furthermore, the effects of target occlusion, introduced by mask patterns positioned at different locations along the optical path, as well as the degradation caused by salt-and-pepper noise, have been systematically analyzed in terms of their impact on inference performance. To the best of our knowledge, this is a first demonstration of a single-layer diffractive architecture delivering such high inference performance, thereby setting a new benchmark in compact, resilient, and high-performance optical computing systems.

\subsection{Network Architecture Design of Differential D\textsuperscript{2}NN}
The conventional architecture of diffractive deep neural networks (D2NNs) is composed of a series of cascaded diffractive modulation layers. Under spatially and temporally coherent illumination, the input information traverses through the diffractive network, generating an intensity distribution on the detection plane. This multi-layer configuration introduces significant optical spatial complexity and is susceptible to alignment errors between modulation layers, which can compromise performance. In contrast, the dual-wavelength differential network proposed in this work employs a simplified single-layer architecture, as shown in Figure in 1 (a) and (b). In terms of the optical setup, the network is comprised of three components: an input layer, a single modulation layer, and an output layer. To ensure that machine learning tasks can be completed normally, the input plane is configured with dimensions of $168\times168$ and a grid size of $8\mu m$, with the distance to the diffractive modulation layer set to 5 cm. The diffractive modulation layer contains a grid of $200\times200$ adjustable neurons, with each neuron having a size of $8\mu m$. The output plane is located 5cm away from the diffractive layer along the axial position. The neurons are optically interconnected through diffraction transmission governed by the Rayleigh-Sommerfeld diffraction, enabling the transmission of optical signals between them. It is worth noting that the parameters of this architecture are meticulously optimized to guarantee sufficient optical interconnectivity, enabling high-fidelity information processing and robust task performance.
\begin{figure}[ht!]
	\centering
	\includegraphics[width=1.0\linewidth]{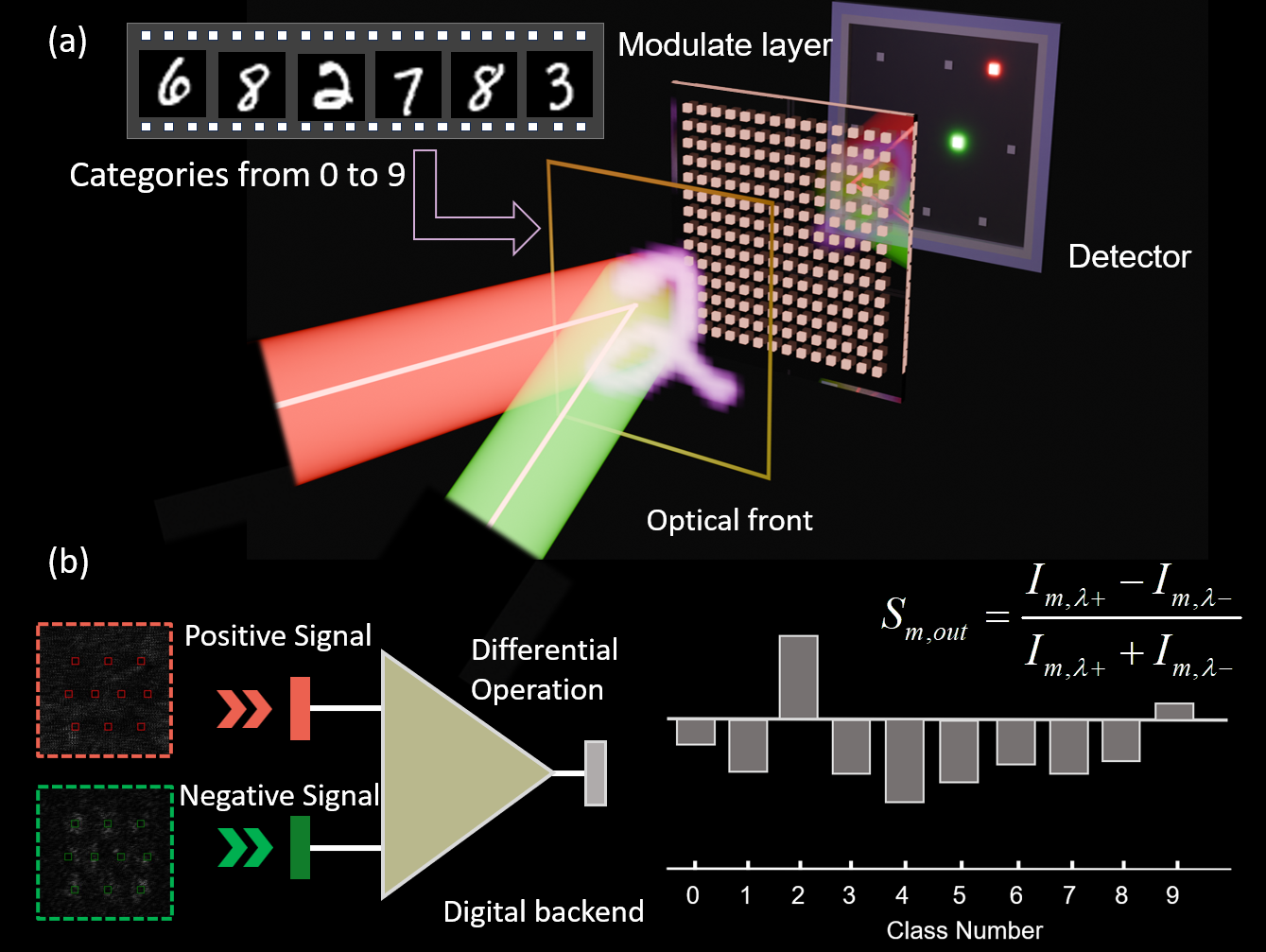}
	\caption{Schematic of the single-layer dual-wavelength differential diffractive neural network (D$^2$NN). (a) Architecture comprising input layer, single diffractive modulation layer with trainable phases, and output detection plane divided into class regions. Dual-wavelength illumination ($\lambda_+$ and $\lambda_-$) enables multiplexing and differential detection for enhanced classification. (b) Principle: Coherent waves diffract through the shared layer, yielding complementary intensities; normalized differential signals determine class via maximum value, eliminating multi-layer misalignments.}
	\label{fig:enter-label}
\end{figure}
Based on the parameters outlined above, the incident targets, including handwritten digits and fashion items, are encoded onto an amplitude-modulated wavefront modulator. These targets are illuminated with coherent light at two distinct wavelengths, ensuring independent wavelength-specific processing, and then passed through the single-layer diffractive network. This design enables diffractive layer to modulate the optical information illuminated by both wavelengths during each task execution, with the information processing for each wavelength sharing identical diffractive modulation parameters. Mathematically, the diffractive network can be modeled as an optical operator $\mathcal{L}$, which transforms the input target $i\left(\lambda\right) $ for each wavelength \( \lambda \). The optical field undergoes modulation and diffraction according to operator $\mathcal{L}$, yielding the output field \( \hat{O}(\lambda) \) at the detection plane:
\begin{equation}
	\hat{O}(\lambda) = \mathcal{L} \left[{i}(\lambda) \right], \quad \lambda \in \{\lambda_+, \lambda_-\},
\end{equation}
Where $\lambda_+$ and $\lambda_-$ represent the distinct illumination wavelengths applied to the input target $i$, each of which is of dimension $N\times N$ and $i\in C^{N^{2}}$ is the vectorized object complex field. After undergoing free-space diffraction propagation and phase modulation at single diffractive layer, the signal light converges at the detection plane, resulting in two wavelength-dependent intensity distributions. The detection plane is partitioned into ten distinct sub-regions, each with a size of 80 µm, where each sub-region is associated with one of the target categories, denoted as \( C = [c_0, c_1, \dots, c_9] \). The sum of optical intensities in the sub-regions corresponding to wavelength \( \lambda_+ = 1064 \, \text{nm} \) is designated as the positive signal, while the sum for \( \lambda_- = 532 \, \text{nm} \) is designated as the negative signal. Specifically, the normalized intensity signals, \( I_{\lambda_+} = [I_{0, \lambda_+}, I_{1, \lambda_+}, \dots, I_{9, \lambda_+}] \) for \( \lambda_+ \), and \( I_{\lambda_-} = [I_{0, \lambda_-}, I_{1, \lambda_-}, \dots, I_{9, \lambda_-}] \) for \( \lambda_- \), represent the positive and negative signals, respectively. Building upon this, the differential score for a given class $c$ is formulated as follows:
\begin{equation}
	S_c = \frac{1}{T} \cdot \frac{\left( \sum_{i=1}^N \left( I_{\lambda_+, i} - I_{\lambda_-, i} \right) \right)}{\left( \sum_{i=1}^N \left( I_{\lambda_+, i} + I_{\lambda_-, i} \right) \right) + \epsilon}
	\label{eq:2}
\end{equation}
The temperature parameter, $T$, serves as a critical hyperparameter in the neural network, modulating the optimization process by controlling the trade-off between exploration and exploitation during training. This mechanism promotes stable convergence and enhances overall model performance. In this work, the temperature hyperparameter was empirically set to 0.1 based on preliminary validation that demonstrated optimal balance in our optical setup. The summation $\sum_{i=1}^{N}$ represents the total sum of light intensities across the pixels in the specified sub-regions, where $N$ denotes the overall number of pixels included within those sub-regions for the computation. The term $\epsilon$ typically set to a small value such as $10^{-6}$, is included in the denominator to ensure numerical stability by preventing division by zero. The normalized differential signal \( S_c \) is represented as a vector \( \mathbf{S}_c = [S_{c,0}, S_{c,2}, \dots, S_{c,9}] \), where each component \( S_{c,i} \) corresponds to the signal difference for a given sub-region under dual-wavelength illumination. The final classification decision is determined by applying the maximum operation $Max\left(\cdot\right)$, to the differential signal vector $\mathbf{S}_c$, identifying the class with the highest differential score.

For each training task, a random subset of 50 grayscale images was selected from both the MNIST and Fashion-MNIST datasets, each consisting of 60,000 training images and 10,000 test images. Spatial information for each object at the input plane was encoded separately into the amplitude channel for optical processing. The original image, with a size of 28 × 28 pixels, was initially upsampled to 168 × 168 pixels using nearest-neighbor interpolation. Subsequently, a zero-padding operation was applied to the input image to ensure it conformed to the required dimensions of the diffractive network. To design the phase parameter of each neuron within the diffractive network, we optimize the single diffractive layer/surface using deep learning. During the training process, the input target propagates forward under two distinct wavelength illuminations, with both illuminations sharing the same of diffractive modulation layer parameters. The optimized diffractive layer comprises tens of thousands of diffractive features (terms as neurons), where the individual phase values of these neurons are adjusted in the training phase through error back-propagation, by minimizing a customized loss function between the one-hot encoding and the predicted differential score.  During each iteration, the diffractive network performed forward propagation on a total of 60,000 samples, utilizing two distinct wavelengths. At the output plane, optical intensity signals corresponding to the two wavelengths $\lambda_{+}$ and $\lambda_{-}$ were measured separately. Specifically, the intensity distributions corresponding to the wavelengths \( \lambda_{+} \) and \( \lambda_{-} \), represented by \( I_{\lambda_{+}} \) and \( I_{\lambda_{-}} \), were separately captured through optical filtering to mitigate cross-talk between the positive and negative signal channels. The intensity values for the two wavelengths in specific sub-regions of the detection plane, corresponding to each class, were used as positive and negative signals to compute the normalized differential score, as defined in Eq.\ref{eq:2}. The loss value, calculated using the soft-max cross-entropy (SCE) loss function, was employed to compute the error gradient, which subsequently updated the phase values of the neurons in the diffractive layers. After training for 50 epochs, the loss value gradually stabilized, indicating convergence and effective learning of the intrinsic features of the datasets.

\section{Results}

\begin{figure}[ht!]
    \centering
    \includegraphics[width=1.0\linewidth]{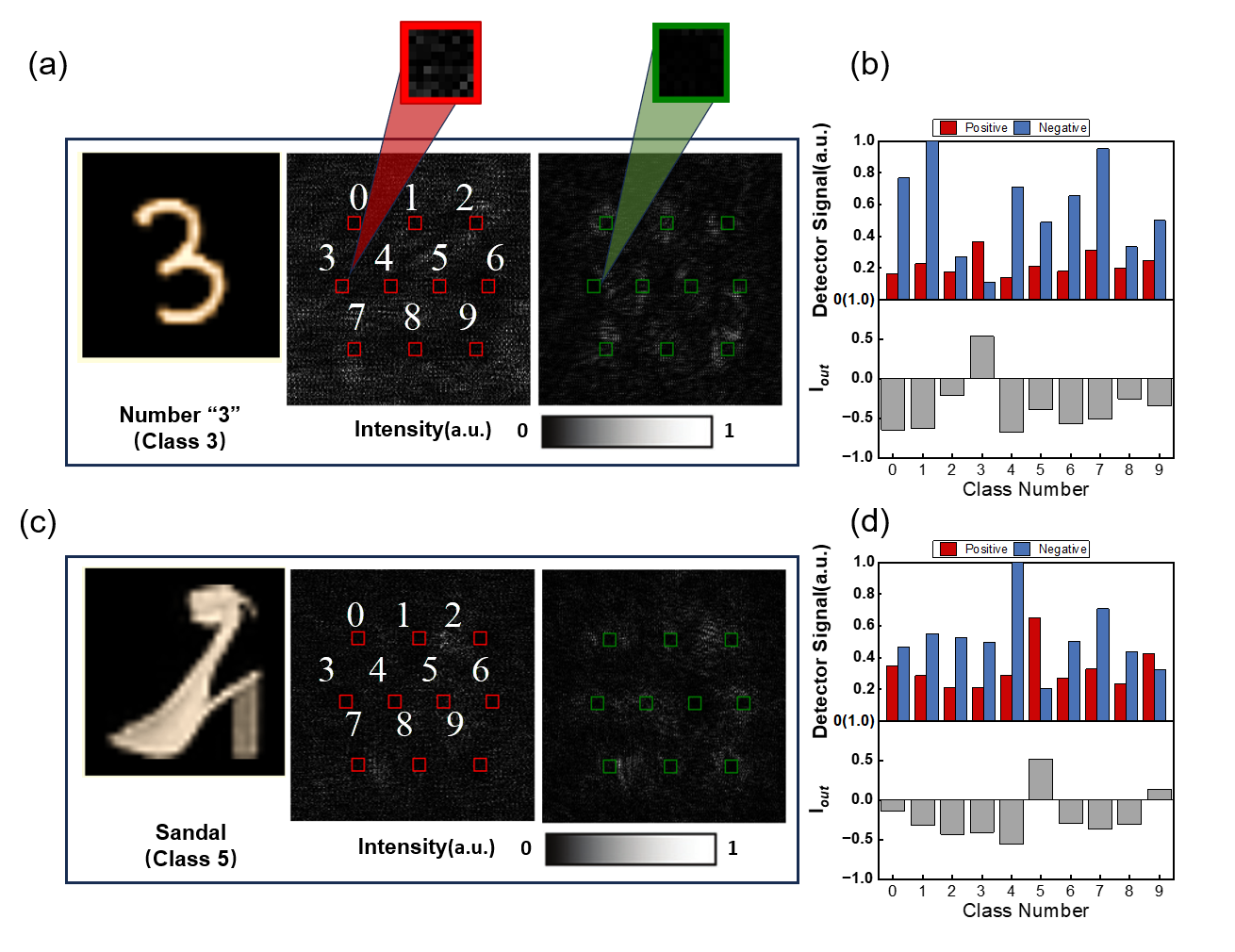}
    \caption{Classification results of the single-layer dual-wavelength differential D$^2$NN. (a, c) Output intensity distributions for MNIST digit 3'' and Fashion-MNIST Sandal'' under $\lambda_+$ (red boxes) and $\lambda_-$ (green boxes), focusing energy in correct classes. (b, d) Normalized differential signals across classes, with maxima indicating accurate predictions. Dual wavelengths capture complementary features, outperforming single-wavelength shallow networks.}
    \label{fig2}
\end{figure}

To demonstrate the superiority of our proposed single-layer diffractive network with dual-wavelength differential, we independently trained two distinct diffractive networks using samples from the MNIST and Fashion-MNIST datasets, respectively. After of deep learning training, which is one time effort, all the parameters of the neurons in the diffractive layer are fixed. To evaluate the performance of diffractive network, we employed test set samples that had not been involved in the training phase. As shown in Fig.\ref{fig2} (a) and (c), the trained diffractive network with dual-wavelength differential successfully identified the correct categories of these objects, utilizing only a single diffractive modulation layer. Specifically, the handwritten digit "3" and the fashion item "Sandal" (category "5") were illuminated with wavelength ($\lambda_+$) and wavelength ($\lambda_-$), respectively, before being processed by the diffraction network. After passing through the trained diffraction networks, intensity distributions were captured on the detection plane. The sub-regions of interest are highlighted by red and green boxes, corresponding to the positive wavelength $\lambda_+$, while the green box highlights the intensity distribution under the negative wavelength $\lambda_-$. Fig.~\ref{fig2} (b) and (d) depict the differential signals for the digit "3" (Class 3) and the sandal (Class 5), respectively. In both cases, the sum of optical intensity corresponding to the positive signal ($\lambda_+$) exhibits a significantly higher intensity than that of the negative signal ($\lambda_-$). The final classification is ascertained by maximizing the differential score $S_c$, as formalized in Eq.~\ref{eq:2}, thereby precisely delineating the target category through normalized and differentiated signals from the dual wavelengths.

In contrast, a conventional single-layer diffractive network under single-wavelength illumination fails to accurately classify the same targets—handwritten digit "2" and fashion item "Sandal", owing to the inherent limitations of its shallow architecture. This limitation stems from the fundamental constraints of the single-layer architecture. With merely one layer comprising $200\times200$ neurons, such a system lacks the depth required to effectively extract and process the intricate features embedded in the input data. On the other hand, by leveraging our proposed dual-wavelength differential scheme within the single layer diffraction network, both the handwritten digit "2" and the fashion item "Sandal" are accurately recognized. The superior performance can be attributed to the synergistic advantages of dual-wavelength illumination, which harnesses intensity signals from two distinct wavelengths to capture complementary spatial frequency information. Specifically, the dual-wavelength differential scheme exploits diffraction intensity signals from two distinct wavelengths, each capturing unique spatial frequency information. This multi-wavelength strategy enables the network to detect wavelength-dependent diffraction patterns, thereby extending its capacity to resolve subtle spatial features that would remain undetected by the traditional single-wavelength regime.

\begin{figure}[ht!]
    \centering
    \includegraphics[width=1.0\linewidth]{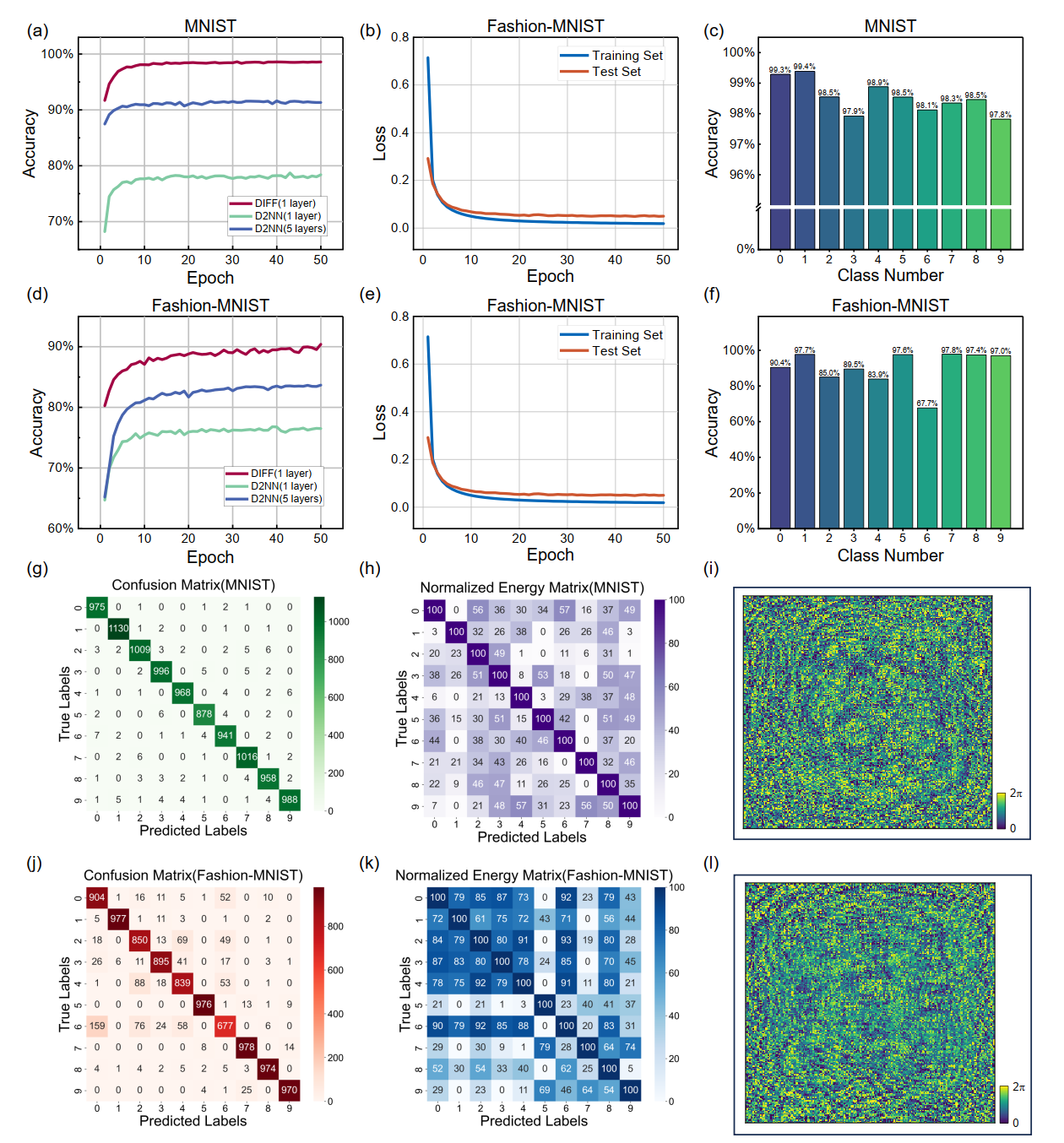}
    \caption{Performance evaluation of the single-layer dual-wavelength differential D$^2$NN. (a, d) Confusion matrices for MNIST and Fashion-MNIST (accuracies: 98.59\% and 90.4\%). (b, e) Normalized energy matrices. (c, f) Class-wise accuracy distributions. (g, j) Accuracy comparisons with single-/five-layer D$^2$NNs, showing superiority with fewer parameters. (h, k) Loss curves. (i, l) Optimized phase distributions (0 to $2\pi$).}
    \label{fig3}
\end{figure}

To quantitatively evaluate the performance of our proposed model, we tested it on 10,000 previously unseen samples and recorded the differential results for each test instance. The performance of our proposed single layer diffractive model demonstrates the significant improvements in classification accuracy compared to traditional diffractive network under the single wave-length illumination, as shown in Fig.~\ref{fig3} (a,b) and (d,e). Specifically, our proposed diffractive network, utilizing a single diffraction modulation layer (200$\times$200 neurons), achieved classification accuracies of 98.59\% and 90.4\% on the handwritten digit and fashion item datasets, respectively. By comparison, the traditional diffractive network under the same network architecture and single-wavelength illumination achieved only 78.39\% accuracy on MNIST and 76.50\% on Fashion-MNIST. Furthermore, even a conventional five-layer cascaded diffractive deep neural network, consisting of 200,000 neurons, attains a maximum classification accuracy of 91.33\% on MNIST and 83.67\% on Fashion-MNIST, respectively. Beyond accuracy gains, our single‑layer design markedly simplifies the optical hardware. Eliminating multiple modulation layers sidesteps the cumulative misalignment errors that typically compromise deeper cascades, thereby enhancing both robustness and operational efficiency. Fig.~\ref{fig3} (i and l) illustrates the phase profile distribution of the single modulation layer (for both MNIST and Fashion-MNIST), revealing phase distributions with distinct local-focus characteristics. The phase values of each neuron are constrained within the range of 0 to $2\pi$, facilitated by the application of the sigmoid activation function $\sigma(x) = \frac{1}{1 + e^{-x}}$.

\begin{figure}
	\centering
	\includegraphics[width=1.0\linewidth]{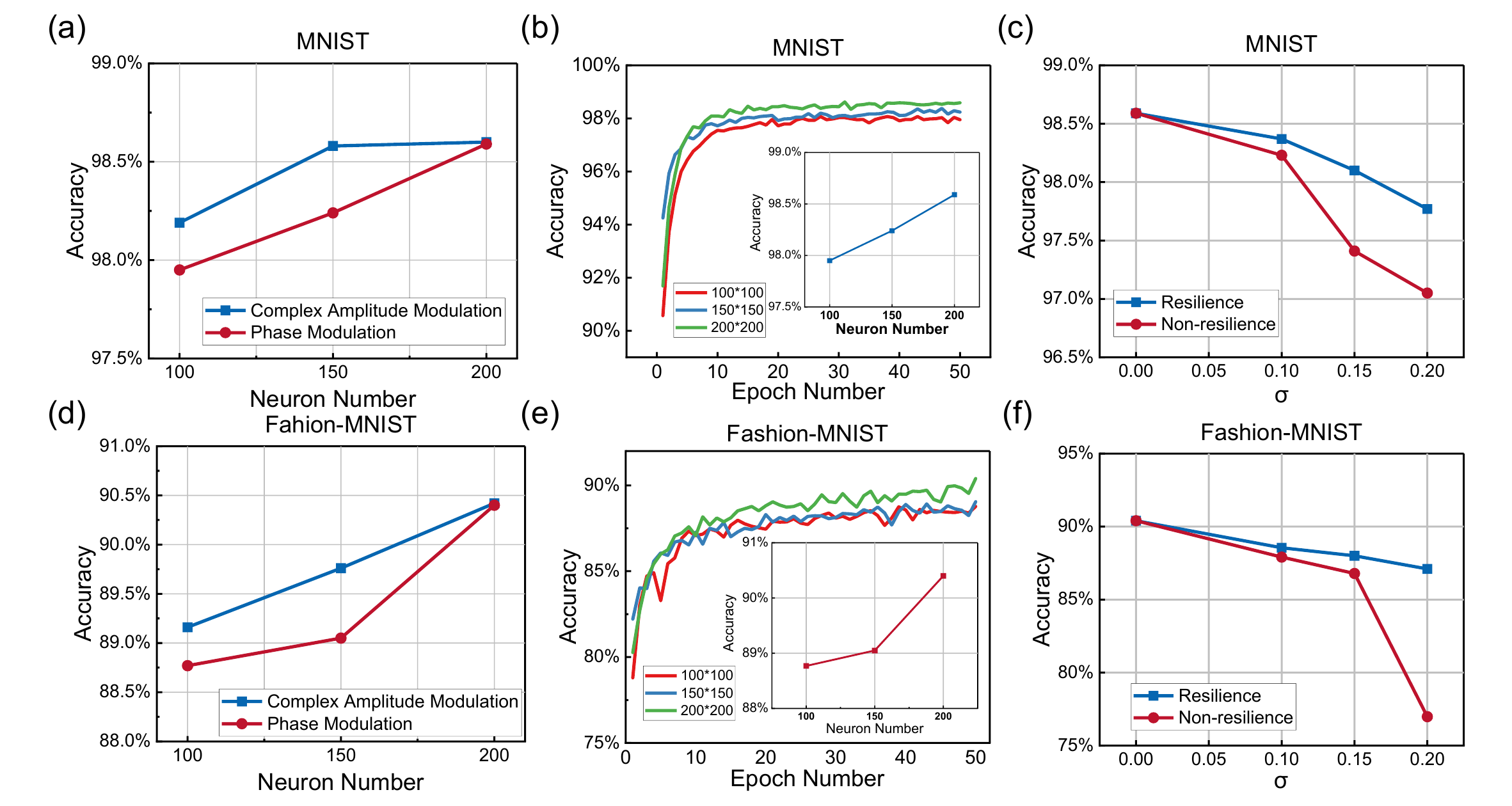}
	\caption{Effects of modulation and resilience in the single-layer dual-wavelength differential D$^2$NN. (a, d) Accuracy vs. neuron count for complex amplitude vs. phase-only modulation, with complex outperforming at low densities. (b, e) Accuracy at reduced parameters (97.95\% and 88.7\% at 10k). (c, f) Resilience to phase errors ($\sigma$): Resilience-trained networks maintain accuracy up to $\sigma=0.2$, mitigating imperfections.}
	\label{fig4}
\end{figure}
 Fig.\ref{fig4} (d) presents a comprehensive evaluation of the classification performance of the proposed single-layer dual-wavelength diffractive neural network under two distinct modulation schemes, namely complex amplitude modulation and phase-only modulation. The results demonstrate that complex amplitude modulation consistently outperforms phase-only modulation as the number of neurons in the diffractive layer increases, achieving higher classification accuracies across both the MNIST and Fashion-MNIST datasets. This performance advantage is particularly pronounced at lower neuron counts (e.g., 100 $\times$ 100 neurons), where the additional degree of freedom provided by amplitude modulation enhances the network’s capacity to encode and process complex spatial frequency information of input objects. Through concurrent modulation of amplitude and phase, the system adeptly encompasses an expansive repertoire of diffraction modalities, thereby bolstering discriminative feature extraction and fortifying classification resilience against perturbations. However, as the neuron count approaches 200$\times$200, the performance gap between the two modulation schemes narrows, with both achieving comparable accuracies (e.g., 98.59\% for MNIST and 90.40\% for Fashion-MNIST under complex amplitude modulation). This convergence suggests that, at higher neuron densities, phase-only modulation can sufficiently exploit the increased spatial resolution to approximate the expressive power of complex amplitude modulation. Meanwhile, Fig.\ref{fig4} (b) and (e) demonstrate that even when the neuron count in the single‑layer diffractive network is reduced to 100$\times$100, it still achieves high classification accuracies of 98\% and 88.9\%, respectively.

Despite employing only a single diffractive modulation layer, the dual-wavelength differential paradigm proposed in our study achieves superior classification accuracy while inherently eliminating the inter-layer mechanical alignment errors that compromise conventional multi-layer diffractive architectures. Nonetheless, inherent imperfections of phase modulate inevitably introduce random phase errors, owing to the limitations in manufacturing precision. To shed more light on the impact of these random phase errors on the performance of diffractive network, which were modeled as uniformly distributed random variables $\omega_{\alpha}(m,n)\sim\mathcal{\alpha\times U}(m,n)$, where $\mathcal{U}(m,n)$ follows the uniform random variable between 0 to $2\pi$, independently generated for each diffractive feature at the trained diffractive layer. The tested results under the presence of random phase errors at the diffractive layers, as shown in Fig.\ref{fig4} (c) and (f) red dotted line. As the random phase error coefficient $\alpha $ of the single-layer diffractive surface increases, the network’s classification accuracy precipitously declines, this because the forward propagation of the object’s optical field diverging from the idea designed phase-mapping transformation. On the other hand, this negative impact caused by random fabrication error can be mitigated considerably. Specifically, we deliberately introduce the random phase errors into the forward model of the diffractive network during the training process, which will force the deep learning– based optimization of the classification tasks to converge to robust solutions against these random phase perturbations at the single diffractive layers. Fig.\ref{fig4} (c) and (f) blue dotted line also report the performance of the resilience-enhanced diffractive network under random phase perturbations. Compared to the baseline architecture, this resilience strategy markedly mitigates the deleterious effects of phase errors, thereby preserving classification accuracy.

\section{Discussion}
\begin{figure}[ht!]
    \centering
    \includegraphics[width=1.0\linewidth]{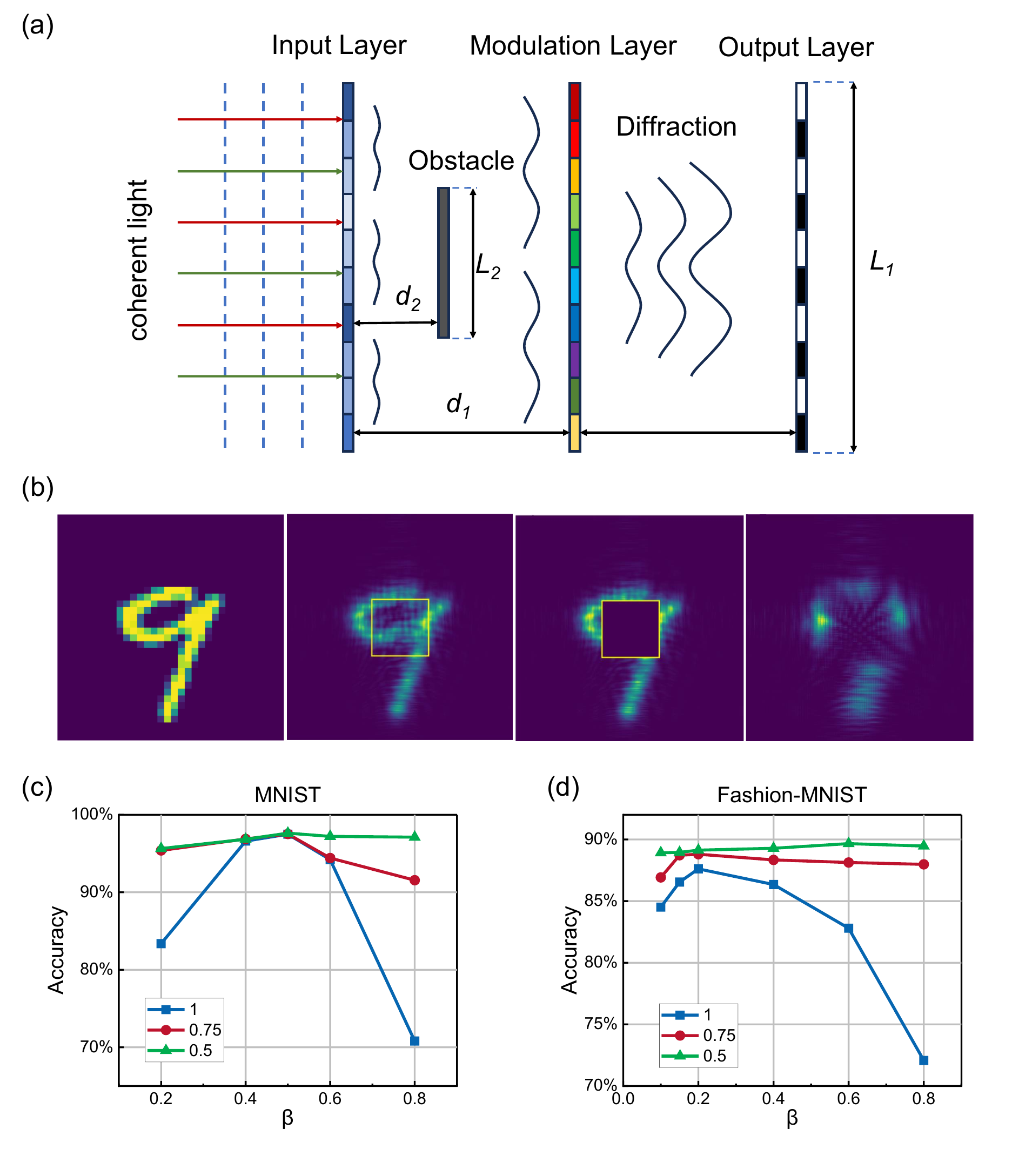}
    \caption{Robustness to occlusions in the single-layer dual-wavelength differential D$^2$NN. (a) Setup with opaque occlusion (width $\epsilon L_1$) between input and modulation layers, using edge diffraction. (b, c) Accuracy vs. position ($\beta$) for varying $\epsilon$ (0.25--0.75) on MNIST/Fashion-MNIST, peaking at 98.1\%/90.2\% for $\epsilon=0.25$, outperforming five-layer D$^2$NNs under severe occlusions via spectral multiplexing.}
    \label{fig5}
\end{figure}

 Due to the loss of feature information, the classification of input objects affected by environmental interference and occlusions are challenging issues for traditional diffractive network. Here, we use a single layer differential diffractive network with dual-wavelength illumination to test the performance of this network under the presence of complex scenarios. For the classification task of samples blocked by occlusions, we further demonstrate an optical architecture for directly recognizing the optical information of interest around a opaque obstruction, where the optical path is either partially or entirely obstructed by opaque occlusions with zero light transmittance, as shown in Fig.\ref*{fig5} (a). Fig.\ref*{fig5} (b) presents the input target, the corresponding intensity distributions before and after the occlusion plane following free-space propagation, and the resulting intensity pattern at the diffractive modulation plane. It can be observed that, despite the presence of an opaque occlusion along the optical path, part of the optical information from the input target is still transmitted to the modulation plane via edge diffraction. For this occlusion-involved recognition scheme, a single diffractive modulation layer is likewise trained in a data-driven manner for effective recognizing the input object, bypassing the fully opaque occlusion positioned between the input plane and modulation layer.
 
To bring more insights into the occlusion width $L_{2}=\varepsilon L_{1}$, where $L_{1}$ denotes the size of the diffractive modulation layer, $\varepsilon$ represents a coefficient, ranging from 0 to 1. Specifically, \(\varepsilon = 0\) corresponds to the absence of occlusion, \(\varepsilon = 1\) indicates complete occlusion, and values between 0 and 1 represent partial occlusion. First, we compare, for various position of opaque occlusions, the performance of trained diffractive network in terms of difference of opaque occlusions size. Specifically, for MNIST dataset, the network achieves a peak classification accuracy of 98.1\% at $\varepsilon \approx 0.5$, significantly outperforming the traditional five-layer D\textsuperscript{2}NN, which yields only 80\% accuracy at the same $\varepsilon$. On the other hand, for the informationally complex Fashion-MNIST dataset, accuracy peaks at $\varepsilon \approx 0.2$. This discrepancy stems fundamentally from the datasets' disparate Fourier spectral compositions. MNIST numerals, primarily comprising low-spatial-frequency contours from broad strokes, preserve perceptual integrity through diffractive edge effects, enabling coherent reconstruction despite significant aperture reduction. In contrast, Fashion-MNIST exemplars, enriched with high-spatial-frequency elements such as intricate textile patterns, necessitate an expansive effective aperture to resolve these fine-scale features, thereby exhibiting heightened susceptibility to occlusion-induced truncation of high-frequency information in the diffraction field. The dual-wavelength architecture, employing wavelengths $\lambda_+$ and $\lambda_-$, generates complementary diffraction patterns, enhancing feature discrimination through differential intensity computation defined in Eq.\ref{eq:2}. This spectral multiplexing ensures robust classification, particularly for MNIST at $\varepsilon \approx 0.5$, where sufficient diffracted light preserves critical low-frequency components.

\begin{figure}[ht!]
    \centering
    \includegraphics[width=1.0\linewidth]{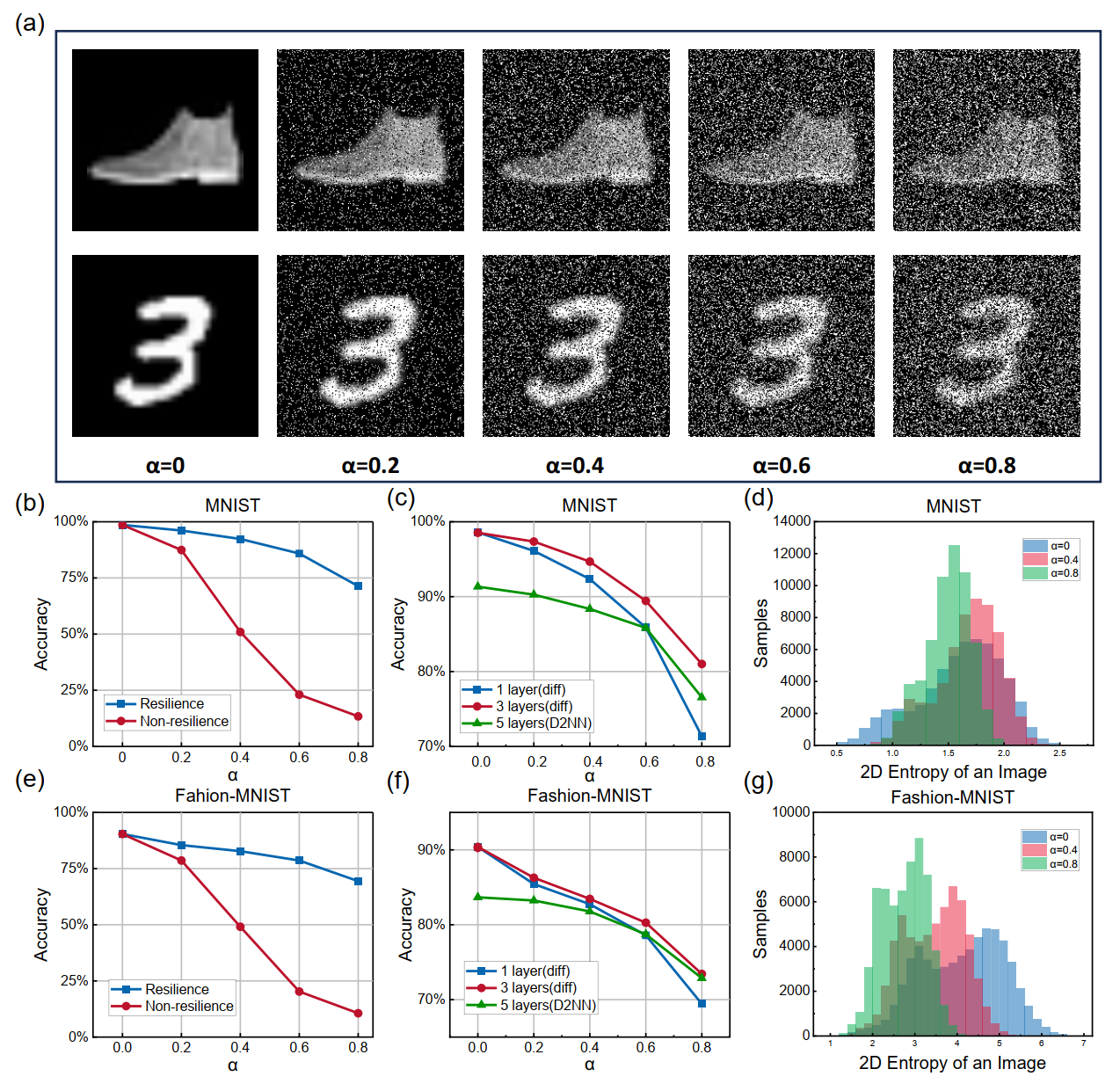}
    \caption{Noise resilience of the single-layer dual-wavelength differential D$^2$NN. (a) Inputs with salt-and-pepper noise ($\alpha=0$--0.8). (b, e) Accuracy vs. $\alpha$ for resilient vs. non-resilient networks, resilient sustaining >50\% at $\alpha=0.8$. (c, f) Performance across depths. (d, g) 2D entropy distributions, shifting with noise and correlating to accuracy drops, highlighting differential efficacy for complex datasets.}
    \label{fig6}
\end{figure}

In addition to opaque obstructions along the optical path, input objects in practical scenarios are frequently corrupted by random diffusers and environmental noise, with salt-and-pepper noise—a form of high-intensity impulse noise—posing a particularly severe challenge by sporadically flipping pixel values to extremes, thereby disrupting feature integrity in optical information processing. While the dual-wavelength differential diffractive neural network (D$^2$NN) excels in feature extraction through spectral-domain multichannel fusion and differential computation, its robustness under such noisy conditions warrants rigorous evaluation. As illustrated in Fig.~\ref{fig6}(a), exemplary images from the MNIST and Fashion-MNIST datasets degrade progressively with increasing noise ratio $\alpha$ (from 0 to 0.8), manifesting as scattered black-and-white artifacts that obscure structural details. For the MNIST dataset, dominated by low-frequency stroke contours, the D$^2$NN mitigates noise-induced degradation by exploiting complementary wavelength-specific diffraction patterns, sustaining high classification accuracy (e.g., ~98.5\% at low $\alpha$). In contrast, the Fashion-MNIST dataset, rich in high-frequency textural elements, exhibits heightened vulnerability, yet the proposed architecture maintains robust performance (e.g., ~90.4\% accuracy) via normalized intensity differentials (as defined in Eq.~\ref{eq:2}), which counteract the non-negativity limitations of single-wavelength systems and enhance discriminative power. Complementing these observations, the 2D entropy distributions in Fig.~\ref{fig6}(d) and (g) provide an information-theoretic perspective, at $\alpha=0$, entropy histograms are compact and low-valued (~1.5–2.0), reflecting ordered image structures conducive to accurate classification; however, as $\alpha$ escalates to 0.8, distributions broaden and shift toward higher entropy (~2.5–3.0), quantifying the injected randomness and elevated complexity that correlate directly with accuracy declines, underscoring how noise amplifies informational disorder and challenges optical feature propagation.

Fig.~\ref{fig6}(b) and (e) further elucidate the impact of salt-and-pepper noise on accuracy versus $\alpha$, contrasting resilient and non-resilient D$^2$NN configurations, while (c) and (f) delineate performance across varying network depths. Non-resilient networks (red curves) suffer abrupt accuracy drops, plummeting to near-zero at high $\alpha$ due to unmitigated noise propagation through diffractive layers. Resilient variants (blue curves), trained with noise-aware strategies, exhibit markedly flatter degradation profiles, preserving ~50\% accuracy at $\alpha=0.8$ by adaptively compensating for pixel corruptions via residual-like mechanisms that stabilize phase and intensity mappings. This resilience is amplified in deeper architectures (up to 5 layers, green curves), where hierarchical diffraction enables contextual feature integration, yielding 20–40\% accuracy gains over shallow counterparts at $\alpha>0.4$, particularly in the more complex Fashion-MNIST. Notably, the single-layer D$^2$NN outperforms traditional five-layer designs (98.5\% vs. 91.33\% on MNIST; 90.4\% vs. 83.67\% on Fashion-MNIST) with over 50\% parameter reduction (40k vs. 200k), attributing this efficiency to minimized inter-layer misalignment and synergistic dual-wavelength fusion. The entropy analysis reinforces these trends, as noise-driven complexity hikes evident in wider histograms—exacerbate recognition challenges in non-resilient or shallow networks, while resilient, depth-optimized D$^2$NNs effectively curtail effective entropy, fostering noise-tolerant optical computing for compact, real-world applications.

\section{Conclusion}
In this work, we proposed a pioneering single-layer dual-wavelength differential diffractive network that addresses critical limitations of conventional single-wavelength multilayer systems. By leveraging coordinated dual-wavelength phase modulation and differential detection within a single diffractive layer, the system achieves high-accuracy classification (98.59\% for MNIST, 90.40\% for Fashion-MNIST) while reducing hardware complexity by over 50\%. The spectral-domain multi-channel information fusion enhances feature extraction capabilities, overcoming the nonnegativity constraint and spectral information loss inherent in traditional designs. Furthermore, the single-layer architecture eliminates inter-layer misalignment errors, a critical bottleneck in multi-layer D$^2$NNs, significantly enhancing scalability and robustness against environmental perturbations such as random phase errors and optical path occlusions. The incorporation of a resilience strategy during training further fortifies the network’s performance, maintaining high accuracy under extreme noise conditions (\(\alpha \leq 0.8\)) and partial occlusions (\(\epsilon \approx 0.5\)). These advancements position our D$^2$NN as a transformative solution for compact, energy-efficient, and noise-resilient optical computing, with broad implications for real-time applications in medical imaging, edge-device pattern recognition, and adaptive signal processing. Future work will explore multi-wavelength extensions and tunable materials to further enhance dynamic reconfigurability, paving the way for next-generation photonic computing platforms.

\begin{acknowledgement}

This work was supported by the National Natural Science Foundation of China (Grant No. 62405296), the Laser Fusion Research Center Funds for Young Talents (Grant No. RCFCZ7-2024-7) and the Postdoctoral Fellowship Program of CPSF under Grant Number (Grant No. GZC20233520).

\end{acknowledgement}
\section{Methods}
\subsection{Diffractive Forward Model for Differential Network}
The proposed single-layer dual-wavelength differential diffractive neural network (D$^2$NN) integrates optical diffraction principles with deep learning optimization. The network architecture consists of an input layer, a single diffractive modulation layer comprising trainable neurons, and an output detection plane. Each neuron in the diffractive layer acts as a secondary wave source according to Huygens' principle, modulating the incident wavefront through phase adjustments.

The optical field propagation is modeled using the Rayleigh-Sommerfeld diffraction equation, implemented digitally via the angular spectrum method for computational efficiency. Two coherent plane waves at distinct wavelengths ($\lambda_+$ and $\lambda_-$) illuminate the amplitude-encoded input image at the $z=0$ plane. The impulse response $h_i^l(x, y, z)$ in the spatial domain is given by:
\begin{equation}
     h_i^l(x, y, z) = \frac{z - z_i}{r^2} \left( \frac{1}{2\pi r} + \frac{1}{j\lambda} \right) \exp\left( j \frac{2\pi r}{\lambda} \right)
\end{equation}

where $h_i^l(x, y, z)$ refers to the light field distribution at the
diffraction neuron with coordinates $(x,y,z)$, which is generated by the $i-th$ neuron with coordinates $\left(x_{i},y_{i},z_{i}\right)$, $r=\sqrt{x^2+y^2+z^2}$, $j=\sqrt{-1}$ refers to the
imaginary unit, the modulated optical field immediately following the diffractive layer located at $z=z_{1}$ was formulated as

\begin{equation}
u_{i}(x,y,z_{1})=t\left(x,y,z_{1}\right)\left[h_{i}(x,y,z_{1}-z_{0})\ast u_{i}(x,y,z_{0})\right]
\end{equation}

where, $t(x,y,z_{1})$ denotes the transmission‑coefficient of each neuron, which is consist of two part: amplitude and phase. For the case of phase only modulation, the value of amplitude was set to be 1 throughout this paper. $u_{i}(x,y,z_{0})$ is the optical field on the input plane $z_{0}$, and “$\ast$” represents the 2‑D convolution operator. After modulation, the field $u_{i}(x,y,z_{1})$ propagates to the detection plane, where the resulting intensity pattern $I_{i}(x,y)$ is measured and can be written as:

\begin{equation}
     I_{i}\left ( x,y \right )=\left |h_{i}\left ( x,y,z-z_{1} \right )\ast u_{i}\left ( x,y,z_{1} \right ) \right |^{2}
\end{equation}

The cross-entropy loss function is employed in this work to optimize the model parameters. The mathematical formulation is expressed as:

\begin{equation}
\mathcal{L} = -\frac{1}{B} \sum_{i=1}^{B} \sum_{m=1}^{C} y^{i}_{m} \log\left(\text{softmax}({I}^{i}_{out,m})\right)
\end{equation}

where \( B \) denotes the batch size, \( C \) represents the number of classes (e.g., \( C=10 \) for the MNIST dataset); \( y^{i}_{m} \in \{0,1\} \) is the one-hot encoded ground-truth label  for the \( i'th \) sample; \( {I}^{i}_{out,m} \) corresponds to the raw model output for the \( i'th \) sample; \( \text{softmax}({I}^{i}_{out,m})= \frac{\exp({I}^{i}_{out,m})}{\sum_{k=1}^{C} \exp({I}^{i}_{out,k})} \) calculates the predicted probability of the \( i'th \) sample belonging to class \( m \).

In this work, salt-and-pepper noise was utilized to validate our model. Salt-and-pepper noise appeared as randomly distributed black and white pixels in an image, creating a noticeable speckled effect. It was introduced to the original image according to the following formula: 
\begin{equation}
 I_{Noise}\left ( x,y \right )=\left\{\begin{matrix}
0,&&\left ( x,y \right )\in Pepper&region\\
1,&&\left ( x,y \right )\in Salt&region\\
I\left ( x,y \right ),&&otherwise
\end{matrix}\right.
\end{equation}
$\alpha$ is set as the ratio of the noise.The areas of the pepper region and salt region are equal.
\subsection{Other Details of Differential D\textsuperscript{2}NN }
Object classification performances of all the models presented in this paper were trained and tested on two widely used datasets: MNIST and Fashion-MNIST. For each dataset, 50,000 samples were used as training data while the remaining 10,000 objects were used for testing. Each original sample is converted into an image with a resolution of 168 × 168 and then zero-padded to a resolution of 200 × 200. The scale of the input object is squeezed proportionally to match the neural layer width reduction.

In each iteration, a batch size of 32 different images from the customized datasets was randomly sampled; correspondingly, 20 differential results at the output plane were used to calculate the loss function. The network was trained using Python (v.3.9.21) and PyTorch (v.2.0.0). The backpropagation algorithm was used to calculate the error between the reconstructed image and the object image. An Adam optimizer was used to update the diffractive layer parameters, the learning rate was set to 0.005, and the decaying rate was set to 0.98 epoch × $10^{-3}$, which is the number of current iterations. 
For the training of our models, we used a desktop computer with an NVIDIA GeForce RTX 3060 graphical processing unit (GPU) and Intel Core (TM) i7-12700H CPU @2.30 GHz and 16 GB of RAM, running the Microsoft Windows 11 operating system.

\bibliography{achemso-demo}

\end{document}